\documentclass[fleqn]{article}
\begin{document}
\def\scri{
\unitlength=1.00mm
\begin{picture}(3.0,2.5)(3.5,3.8)
\thinlines
\put(4.9,5.12){\makebox(0,0)[cc]{$\cal J$}}
\bezier{20}(6.27,5.87)(3.93,4.60)(4.23,5.73)
\end{picture}}
\newcommand{\PACS}{04.20.Gz, 04.20.Jb, 04.70Bw}
\newcommand{\shorttitle}
{D. Brill, 2+1-dimensional black holes} 
\title{2+1-dimensional black holes with momentum\\ and angular momentum}
\author{D. Brill\\ 
{\small Department of Physics, University of Maryland}\\ {\small College Park,  
  MD 20782, USA}\\
{\small email:{\tt Brill@physics.umd.edu}}} 
\maketitle
\begin{abstract}
Exact solutions of Einstein's equations in 2+1-dimensional anti-de Sitter 
space containing any number of black holes are described. In addition to
the black holes these spacetimes can possess ``internal'' structure.
Accordingly the generic spacetime of this type depends on a large number of
parameters. Half of these can be taken as mass parameters, and the
rest as the conjugate (angular) momenta. The time development and horizon
structure of some of these spacetimes are sketched.
\end{abstract}

\section{Introduction}
\label{intro}
The discovery \cite{BTZ} that 2+1-dimensional sourcefree Einstein gravity 
with a negative cosmological constant admits black hole spacetime was 
initially surprising because this theory does not admit local gravitational 
degrees of freedom: If the Ricci tensor is constant so is the Riemann tensor, 
spacetime has constant negative curvature, and is therefore locally anti-de 
Sitter (adS). Subsequently multi-black-hole configurations were found and 
classified \cite{DB,DBS}, but only in the time-symmetric context, when the 
gravitational momentum variables (extrinsic curvature) vanish. After a condensed 
review of these time-symmetric spacetimes in Section 2, we discuss in section 
3 the more general case when the momenta do not vanish. Our conclusions are 
summarized in Section 4.

\section{Time-symmetric multi-black-holes}
\label{ts}
\subsection{Initial States}

By definition, time-symmetric geometries possess a spacelike surface $S$ such
that reflection about this surface is an isometry. For multi-black-hole 
solutions this surface is Cauchy, so it suffices to classify the states at 
the moment of time-symmetry, on the two-dimensional spacelike $S$ whose 
extrinsic curvature vanishes. Because the three-dimensional spacetime has
constant curvature $\Lambda < 0$ (where $\Lambda$ is the cosmological constant;
usually replaced by $\ell^2 = -1/ \Lambda$),
the instrinsic geometry of $S$ is also one of constant negative curvature. 
Any such space can be put together out of pieces\footnote{One standard 
construction uses a single piece, a fundamental domain of the discrete 
group $\cal G$ of isometries that specifies which parts of the domain's 
boundary are to be identified, so that $S = H^2/\cal G$. We will however
use an equivalent but somewhat different description. For details see
\cite{BP}.} of its universal
covering space, the simply-connected two-dimensional space of constant 
negative curvature, $H^2$.

The space $H^2$ is conveniently represented as the {\it Poincar\'e disk},
the region $r < \ell$ of the plane with polar coordinates $(r, \theta)$ and
with the metric
\begin{equation}
   ds^2 = {4\over \left(1-{r^2\over \ell^2}\right)^2}\left(dr^2 + r^2 d\theta^2\right)
\label{pd}
\end{equation}
The map between $H^2$ and the plane of polar 
coordinates $(r, \theta)$ is an equal-angle (conformal) map in which 
geodesics are represented as arcs of Euclidean circles normal to the 
``limit circle'' $r = \ell$ (Fig.\ 1).

\begin{figure}
\unitlength 0.80mm
\linethickness{0.4pt}
\begin{picture}(163.61,47.73)(5,0)
\bezier{100}(9.80,25.00)(9.80,36.28)(19.90,42.68)
\bezier{100}(19.90,42.68)(30.00,47.73)(40.10,42.68)
\bezier{100}(40.10,42.68)(50.20,36.28)(50.20,25.00)
\bezier{100}(9.80,25.00)(9.80,13.72)(19.90,7.32)
\bezier{100}(19.90,7.32)(30.00,2.27)(40.10,7.32)
\bezier{100}(40.10,7.32)(50.20,13.72)(50.20,25.00)
\put(10.00,25.00){\line(1,0){40.00}}
\put(30.00,5.00){\line(0,1){40.00}}
\put(15.50,10.50){\line(1,1){29.00}}
\put(44.50,10.50){\line(-1,1){29.00}}
\bezier{100}(69.80,25.00)(69.80,36.28)(79.90,42.68)
\bezier{100}(79.90,42.68)(90.00,47.73)(100.10,42.68)
\bezier{100}(100.10,42.68)(110.20,36.28)(110.20,25.00)
\bezier{100}(69.80,25.00)(69.80,13.72)(79.90,7.32)
\bezier{100}(79.90,7.32)(90.00,2.27)(100.10,7.32)
\bezier{100}(100.10,7.32)(110.20,13.72)(110.20,25.00)
\bezier{100}(21.12,25.00)(21.12,29.92)(25.53,32.71)
\bezier{100}(25.53,32.71)(29.94,34.92)(34.34,32.71)
\bezier{100}(34.34,32.71)(38.75,29.92)(38.75,25.00)
\bezier{100}(21.12,25.00)(21.12,20.08)(25.53,17.29)
\bezier{100}(25.53,17.29)(29.94,15.08)(34.34,17.29)
\bezier{100}(34.34,17.29)(38.75,20.08)(38.75,25.00)
\bezier{100}(131.18,9.21)(131.18,6.19)(139.08,4.48)
\bezier{100}(139.08,4.48)(147.00,3.12)(154.92,4.48)
\bezier{100}(154.92,4.48)(162.82,6.19)(162.82,9.21)
\bezier{100}(14.93,25.00)(14.93,33.42)(22.46,38.19)
\bezier{100}(22.46,38.19)(30.00,41.96)(37.54,38.19)
\bezier{100}(37.54,38.19)(45.07,33.42)(45.07,25.00)
\bezier{100}(14.93,25.00)(14.93,16.58)(22.46,11.81)
\bezier{100}(22.46,11.81)(30.00,8.04)(37.54,11.81)
\bezier{100}(37.54,11.81)(45.07,16.58)(45.07,25.00)
\thicklines
\bezier{156}(102.49,41.03)(91.21,25.13)(102.74,8.97)
\bezier{156}(77.51,41.03)(88.79,25.13)(77.26,8.97)
\thinlines
\bezier{48}(108.36,33.91)(103.00,30.29)(103.00,24.93)
\bezier{48}(108.36,16.09)(103.00,19.71)(103.00,25.07)
\put(110.00,25.00){\oval(4.00,4.0)[l]}
\bezier{48}(71.64,33.91)(77.00,30.29)(77.00,24.93)
\bezier{48}(71.64,16.09)(77.00,19.71)(77.00,25.07)
\put(70.00,25.00){\oval(4.00,4.0)[r]}
\put(90.00,5.00){\line(0,1){40.00}}
\put(70.00,25.00){\line(1,0){40.00}}
\bezier{124}(69.67,24.93)(74.01,10.29)(89.96,8.99)
\bezier{124}(110.33,24.93)(105.99,10.29)(90.04,8.99)
\bezier{216}(69.6,25.00)(89.96,6.96)(110.30,25.00)
\bezier{124}(69.67,25.07)(74.01,40)(89.96,40)
\bezier{124}(110.33,25.07)(105.99,40)(90.04,40)
\bezier{216}(69.6,25.00)(89.96,43.04)(110.30,25.00)
\put(92.00,25.00){\makebox(0,0)[cc]{1}}
\put(93.00,34.00){\makebox(0,0)[cc]{1.5}}
\put(94.00,40.00){\makebox(0,0)[cc]{2}}
\put(90.00,46.00){\makebox(0,0)[cb]{0}}
\put(103.00,41.00){\makebox(0,0)[lb]{$\pi/2$}}
\put(109.00,34.00){\makebox(0,0)[lc]{$\pi$}}
\put(111.00,26.50){\makebox(0,0)[lb]{$2\pi$}}
\bezier{100}(11.68,25.00)(11.68,35.23)(20.84,41.03)
\bezier{100}(20.84,41.03)(30.00,45.61)(39.16,41.03)
\bezier{100}(39.16,41.03)(48.32,35.23)(48.32,25.00)
\bezier{100}(11.68,25.00)(11.68,14.77)(20.84,8.97)
\bezier{100}(20.84,8.97)(30.00,4.39)(39.16,8.97)
\bezier{100}(39.16,8.97)(48.32,14.77)(48.32,25.00)
\bezier{128}(131.33,40.67)(143.08,37.54)(142.30,25.00)
\bezier{132}(142.30,25.00)(141.52,9.33)(131.33,9.33)
\bezier{128}(162.67,40.67)(150.92,37.54)(151.70,25.00)
\bezier{132}(151.70,25.00)(152.48,9.33)(162.67,9.33)
\bezier{100}(130.39,41.45)(130.39,42.98)(138.69,43.85)
\bezier{100}(138.69,43.85)(147.00,44.53)(155.31,43.85)
\bezier{100}(155.31,43.85)(163.61,42.98)(163.61,41.45)
\bezier{100}(130.39,41.45)(130.39,39.93)(138.69,39.06)
\bezier{100}(138.69,39.06)(147.00,38.38)(155.31,39.06)
\bezier{100}(155.31,39.06)(163.61,39.93)(163.61,41.45)
\put(30.00,0.00){\makebox(0,0)[cc]{(a)}}
\put(90.00,0.00){\makebox(0,0)[cc]{(b)}}
\put(147.00,0.00){\makebox(0,0)[cc]{(c)}}
\put(111.00,25.00){\makebox(0,0)[lc]{$\infty$}}
\put(77.00,41.00){\makebox(0,0)[rb]{$-\pi/2$}}
\put(96.00,44.00){\makebox(0,0)[cc]{$\infty$}}
\put(38.00,28.00){\makebox(0,0)[cc]{$1$}}
\put(44.00,30.00){\makebox(0,0)[cc]{$2$}}
\put(47.00,31.00){\makebox(0,0)[cc]{$3$}}
\put(49.00,32.00){\makebox(0,0)[lc]{$\infty$}}
\end{picture}

 \caption{Representations of initial geometries on a surface of time-symmetry
for black hole spacetimes. (a) Coordinates $(r,\theta)$ on the Poincar\'e 
disk. On this scale and in the metric (\ref{pd}) the circles have the radii 
shown, in units of $\ell$. (b) Coordinates of the BTZ black hole (\ref{BTZ}) 
on the Poincar\'e disk. The geodesics on which $\phi$ has the constant values
shown (for $m = 0.25$) are represented by arcs of Euclidean circles 
that meet the limit circle orthogonally. The horizontal line $r/m = 1$ is likewise 
a geodesic. Other curves, on which $r/m$ has the constant values shown,
are equidistant from this geodesic and have constant, non-vanishing acceleration.
When two copies of the heavily outlined strip, in which $\phi$ changes by $\pi$, 
are superimposed and glued together at the geodesic edges, we obtain the initial
state of the BTZ metric (\ref{BTZ}), shown schematically in (c) as a surface 
in 3-space. This ``pseudosphere'' surface cannot be embedded in Euclidean
space in its entirety. The space itself continues to infinite distance on
both the top and bottom sheet.}
\end{figure}
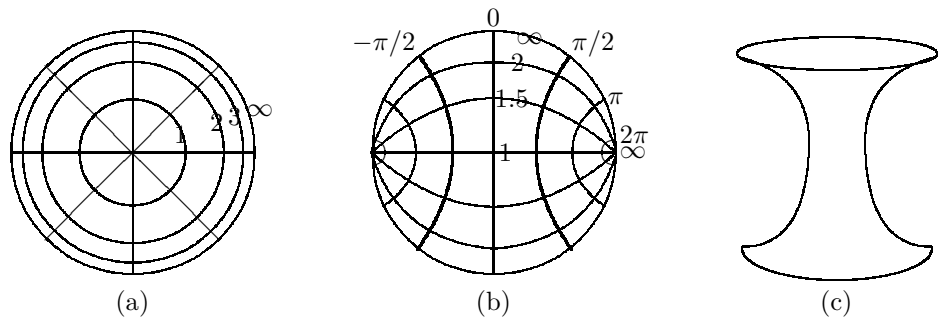

The basic time-symmetric ``single" black hole is that due to Ba\~nados,
Teitelboim and Zanelli (BTZ) \cite{BTZ}, with metric
\begin{equation}
   ds^2 = -\left({\rho^2\over \ell^2}-m \right) dt^2 + 
\left({\rho^2\over \ell^2}-m \right)^{-1}d\rho^2 + \rho^2 d\phi^2.
\label{BTZ}
\end{equation}
Putting $m=1=\ell$ for simplicity we find the coordinates $(\rho,\phi)$ of 
the space part of (\ref{BTZ}) to be related to the $(r,\theta)$ of (\ref{pd}) by
\begin{equation}
 r^2 = {\rho \cosh\phi - 1\over \rho \cosh\phi + 1} \qquad
\cos\theta = \sqrt{\rho^2 - 1\over \rho^2\cosh^2\phi -1}\; ;
\label{BTZP}
\end{equation}
a polar coordinate plot of (\ref{BTZP}) as in Fig.\ 1b shows that the 
coordinates $\rho, \phi$ cover the Poincar\'e disk if $\phi$ is given
an infinite range. But the coordinate $\phi$
of (\ref{BTZ}) is intended to have the usual range $2\pi$ of a polar angle.
Fig. 1b shows in heavy outline a strip of half this size. We obtain the BTZ 
geometry by laying a second, identical copy of this strip on top and sewing 
the edges together (Fig. 1c).

Multi-black-hole initial states can be constructed in an analogous way:
gluing regions together by boundary geodesics of equal length makes a smooth
union since the extrinsic curvature of a 2D geodesic vanishes. Any corners
of the boundary should be $90^\circ$, so that a regular neighborhood is created
when four corners are glued together. Figure 2a shows an example. An identical
copy is to be glued along the heavily drawn boundaries. The lightly
drawn boundaries (with arrows) then become geodesic circles, and these are 
glued to each other.
The result has the topology of a doubly punctured torus shown in Fig.\ 2b.
Each puncture flares out to infinity, and in such a region it is isometric
to an exterior region ($\rho > \ell\sqrt{m}$) of the BTZ initial state. One
may describe this geometry as the initial state of two black holes that are
joined through a common internal torus.   

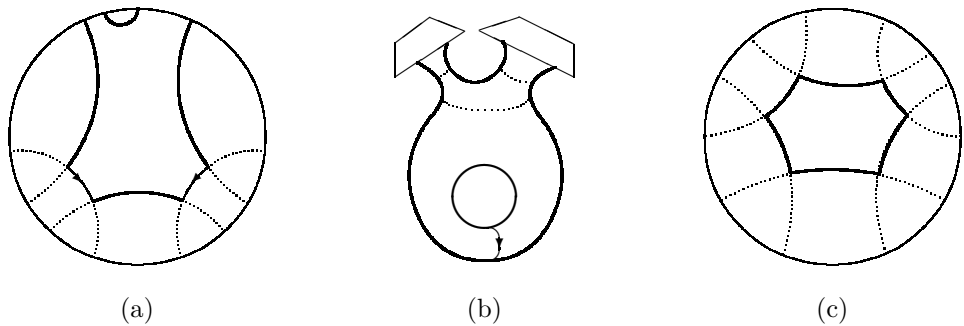
\begin{figure}
\unitlength 0.66mm
\linethickness{0.4pt}
\begin{picture}(215.79,64.01)(21,15)
\bezier{184}(24.21,50.00)(24.21,64.39)(37.11,72.56)
\bezier{180}(37.11,72.56)(50.00,79.01)(62.89,72.56)
\bezier{184}(62.89,72.56)(75.79,64.39)(75.79,50.00)
\bezier{184}(24.21,50.00)(24.21,35.61)(37.11,27.44)
\bezier{180}(37.11,27.44)(50.00,20.99)(62.89,27.44)
\bezier{184}(62.89,27.44)(75.79,35.61)(75.79,50.00)
\bezier{12}(32.53,31.19)(35.99,34.83)(40.98,36.95)
\thicklines
\bezier{160}(64.20,43.86)(54.03,57.87)(60.75,73.42)
\bezier{88}(40.98,36.95)(50.00,40.59)(59.02,37.14)
\bezier{160}(35.80,43.86)(45.97,57.87)(39.25,73.42)
\bezier{16}(43.28,74.77)(44.05,73.04)(45.59,72.65)
\bezier{16}(45.59,72.65)(47.12,72.27)(48.85,73.42)
\bezier{12}(48.85,73.42)(49.81,74.19)(50.00,75.73)
\thinlines
\bezier{13}(59.02,37.14)(64.40,34.83)(67.47,30.81)
\bezier{14}(58.83,25.81)(56.91,31.38)(59.02,36.95)
\bezier{40}(59.02,36.95)(60.75,41.17)(64.20,43.86)
\bezier{15}(64.20,43.86)(69.19,47.89)(75.73,47.12)
\bezier{13}(72.08,36.56)(67.28,39.82)(64.20,43.86)
\bezier{14}(41.17,25.81)(43.09,31.38)(40.98,36.95)
\bezier{40}(40.98,36.95)(39.25,41.17)(35.80,43.86)
\bezier{15}(35.80,43.86)(30.81,47.89)(24.27,47.12)
\bezier{13}(27.92,36.56)(32.72,39.82)(35.80,43.86)
\put(38.68,40.98){\vector(1,-1){0.38}}
\put(61.32,40.98){\vector(-1,-1){0.38}}
\put(50.00,15.00){\makebox(0,0)[cc]{(a)}}
\thicklines
\put(120.00,38.00){\circle{12.00}}
\bezier{84}(120.00,25.00)(130.00,25.00)(134.00,35.00)
\bezier{88}(134.00,35.00)(137.88,45.27)(131.00,54.00)
\bezier{72}(131.00,54.00)(125.76,61.03)(134.00,64.00)
\bezier{84}(120.00,25.00)(110.00,25.00)(106.00,35.00)
\bezier{88}(106.00,35.00)(102.12,45.27)(109.00,54.00)
\bezier{32}(123.51,69.11)(125.10,66.41)(122.24,62.92)
\bezier{44}(122.24,62.92)(117.95,59.11)(113.67,62.76)
\bezier{28}(113.67,62.76)(110.97,65.78)(112.24,68.63)
\bezier{72}(109.00,54.00)(114.78,60.54)(106.37,64.83)
\thinlines
\put(138.00,62.00){\line(-2,1){19.00}}
\put(119.00,71.50){\line(3,1){8.00}}
\put(127.00,74.17){\line(2,-1){11.00}}
\put(138.00,68.67){\line(0,-1){6.67}}
\put(102.00,62.00){\line(3,2){14.00}}
\put(116.00,71.33){\line(-5,2){7.00}}
\put(109.00,74.13){\line(-4,-3){7.00}}
\put(102.00,68.88){\line(0,-1){6.88}}
\put(120.50,28.33){\oval(5.00,6.70)[r]}
\put(123.03,27.68){\vector(0,-1){0.32}}
\put(120.00,15.00){\makebox(0,0)[cc]{(b)}}
\bezier{4}(110.33,62.17)(112.17,62.33)(112.67,63.83)
\bezier{8}(123.00,64.00)(124.50,61.17)(129.33,60.83)
\bezier{12}(111.33,58.33)(111.33,55.50)(120.00,55.33)
\bezier{12}(120.00,55.33)(128.50,55.17)(129.00,58.33)
\bezier{184}(164.21,50.00)(164.21,64.39)(177.11,72.56)
\bezier{180}(177.11,72.56)(190.00,79.01)(202.89,72.56)
\bezier{184}(202.89,72.56)(215.79,64.39)(215.79,50.00)
\bezier{184}(164.21,50.00)(164.21,35.61)(177.11,27.44)
\bezier{180}(177.11,27.44)(190.00,20.99)(202.89,27.44)
\bezier{184}(202.89,27.44)(215.79,35.61)(215.79,50.00)
\bezier{51}(167.67,37.24)(190.05,49.62)(212.43,37.24)
\bezier{47}(200.76,26.52)(193.38,49.86)(213.62,60.81)
\bezier{12}(215.76,50.10)(210.29,49.86)(205.05,54.14)
\bezier{13}(200.05,61.52)(197.90,67.71)(200.52,73.67)
\bezier{49}(209.57,66.76)(190.05,52.24)(173.38,69.86)
\bezier{44}(184.57,75.10)(186.95,53.67)(164.57,50.10)
\bezier{48}(167.19,62.24)(186.95,49.14)(180.05,26.29)
\put(190.05,15.10){\makebox(0,0)[cc]{(c)}}
\thicklines
\bezier{52}(176.48,54.14)(180.29,48.90)(181.48,42.71)
\bezier{72}(200.05,61.29)(191.48,58.90)(183.14,62.24)
\bezier{56}(199.33,42.48)(200.76,49.38)(205.05,54.38)
\bezier{72}(181.48,42.71)(190.29,44.14)(199.33,42.48)
\bezier{36}(205.05,54.38)(201.48,57.48)(200.05,61.29)
\bezier{44}(183.14,62.24)(180.76,57.24)(176.48,54.14)
\end{picture}

\caption{Construction of initial states of multi-black-hole geometries.
(a) The solid black lines are geodesics in the Poincar\'e disk. They
bound a region that has two ends at infinity. (b) Schematic representation
of the geometry obtained by gluing two copies of the region in (a) along
its heavily drawn boundaries, and then identifying the boundaries that
have arrows. The trapezoids indicate the infinite, asymptotically adS
ends. The dotted curves and the curve with the arrow are minimal geodesics
that cut this figure into two semi-infinite ``flares'' and two internal
``cores.'' (c) The general core is obtained by gluing two 
right-angle, geodesic hexagons (one of which is shown) along alternate
sides.}
\end{figure}

In the internal region of a general time-symmetric multi-black-hole 
inital state there
are a number of minimal, homotopically inequivalent, closed geodesics
(the curve with arrow and the dotted curves in Fig.\ 2b). Cutting the
surface along these geodesics decomposes it into ``flares'' and
trousers-shaped ``cores.'' The general core geometry is obtained from
two geodesic hexagons as in Fig.\ 2c, and it is determined by three 
parameters, which can be taken to be the circumferences of the trousers'
legs and waist.

When assembling the general surface\footnote{We confine attention to orientable
geometries; non-orientable ones have a double covering that is orientable.}
out of cores and flares, we have to match the circumferences
of the geodesics at the ``seams,'' but we can join them with an arbitrary
``twist'' (a rotation along the circles). When joining a flare to a 
core, the twist extends to an isometry $\phi\rightarrow\phi+$const of 
the exterior BTZ geometry and produces no new geometry; but a twist
between two cores generally changes the geometry, so at each of those seams
we loose one circumference parameter and gain one twist parameter, with no
net change in the total number of parameters. Thus there are three 
parameters for each core component. If the surface has genus $g$ and
$k$ exteriors, there are $2g-2+k$ cores, and the number of parameters
is $6g-6+3k$. If $g>1$ we can have $k=0$, a finite, closed ``universe.''

The hexagon constituents (Fig.\ 2c) of cores and corresponding infinite
2-gon components of flares are possible coordinate neighborhoods in which
the metric can be given a standard form, such as (\ref{pd}). The
coordinate transformations at the boundary, analogous to (\ref{BTZP}),
increase in number and complexity with $g$ and $k$, but are well-defined
when the gluing scheme is given by a figure like Fig.\ 2a, and the $6g-6+3k$
parameters are specified. In this sense our figures (if labeled with the parameters)
are similar to Feynman diagrams, representing well-defined mathematical 
expressions.

\subsection{Time Development of a black hole}
If we extend the metric of (\ref{BTZ}) to an infinite range of $\phi$, we
obtain a coordinate description of adS spacetime analogous to Rindler coordinates 
in Minkowski spacetime. Figure 3a shows the coordinates of (\ref{BTZ}) on the 
$\phi=0$ section of adS spacetime as
embedded in 2+1-dimensional flat space, as well as their continuation
in the usual way to $\rho<\ell\sqrt{m}$.
Here the horizontal axis is spacelike and planes perpendicular to it are timelike. 
The numbers label the values of the coordinate $t$.

\begin{figure}
\unitlength 0.800mm
\linethickness{0.4pt}
\begin{picture}(122.87,72.61)(-18,3)
\thicklines
\bezier{96}(9.80,66.30)(22.50,55.00)(31.00,55.00)
\bezier{96}(51.00,65.00)(39.50,55.00)(31.00,55.00)
\bezier{96}(9.80,14.70)(22.50,25.00)(31.00,25.00)
\bezier{96}(51.00,15.00)(39.50,25.00)(31.00,25.00)
\bezier{100}(46.09,40.00)(46.09,54.39)(49.55,62.56)
\bezier{100}(49.55,62.56)(53.00,69.01)(56.45,62.56)
\bezier{100}(56.45,62.56)(59.91,54.39)(59.91,40.00)
\bezier{100}(46.09,40.00)(46.09,25.61)(49.55,17.44)
\bezier{100}(49.55,17.44)(53.00,10.99)(56.45,17.44)
\bezier{100}(56.45,17.44)(59.91,25.61)(59.91,40.00)
\bezier{100}(9.80,66.27)(4.88,66.27)(2.09,53.37)
\bezier{100}(2.09,53.37)(-0.11,40.48)(2.09,27.59)
\bezier{100}(2.09,27.59)(4.88,14.69)(9.80,14.69)
\thinlines
\put(28.00,40.00){\line(5,-3){19.00}}
\bezier{64}(30.00,25.00)(28.00,31.54)(28.00,40.00)
\bezier{64}(28.00,40.00)(28.00,50.26)(30.00,55.00)
\bezier{88}(19.00,21.00)(18.50,25.50)(28.00,40.00)
\bezier{84}(28.00,40.00)(36.00,26.17)(39.33,23.00)
\put(28.00,40.00){\line(5,-3){18.67}}
\bezier{72}(28.00,40.00)(39.00,39.33)(46.00,37.83)
\bezier{88}(28.00,40.00)(41.33,48.00)(46.83,52.00)
\put(5.00,63.00){\line(1,-1){44.00}}
\put(5.72,17.78){\line(1,1){43.06}}
\bezier{108}(7.00,17.00)(22.50,32.00)(28.00,32.00)
\bezier{108}(49.00,17.00)(33.50,32.00)(28.00,32.00)
\bezier{92}(48.00,22.00)(36.00,35.00)(36.00,40.00)
\bezier{92}(36.00,40.00)(36.00,45.00)(48.00,58.00)
\bezier{108}(8.00,15.00)(21.50,28.80)(29.00,28.00)
\bezier{92}(29.00,28.00)(33.50,28.00)(48.00,18.00)
\bezier{108}(46.67,51.67)(39.00,38.67)(46.67,29.00)
\put(47.00,38.00){\makebox(0,0)[lc]{$t=0$}}
\put(48.00,52.00){\makebox(0,0)[lc]{1}}
\put(49.33,60.50){\makebox(0,0)[lc]{$\infty$}}
\put(47.67,28.83){\makebox(0,0)[lc]{-1}}
\put(49.17,19.33){\makebox(0,0)[lc]{-$\infty$}}
\put(39.00,22.00){\makebox(0,0)[ct]{-1}}
\put(19.50,20.50){\makebox(0,0)[ct]{1}}
\put(5.00,17.00){\makebox(0,0)[rc]{$\infty$}}
\put(49.00,42.00){\vector(-4,-1){6.00}}
\put(49.00,42.00){\makebox(0,0)[lc]{$\rho=$const}}
\put(30.00,24.50){\makebox(0,0)[ct]{0}}
\put(23.00,40.00){\makebox(0,0)[rc]{O}}
\put(23.00,40.00){\vector(1,0){4.00}}
\bezier{40}(116.60,14.31)(116.60,11.06)(108.76,9.43)
\bezier{40}(108.76,9.43)(100.93,7.81)(93.09,9.43)
\bezier{40}(93.09,9.43)(85.26,11.06)(85.26,14.31)
\bezier{40}(116.60,42.15)(116.60,39.47)(108.76,38.14)
\bezier{40}(108.76,38.14)(100.93,36.81)(93.09,38.14)
\bezier{40}(93.09,38.14)(85.26,39.47)(85.26,42.15)
\bezier{16}(105.63,37.75)(104.67,38.56)(108,40)
\bezier{22}(108,40)(111.42,40.5)(113.46,39.37)
\bezier{16}(96.23,46.54)(97.18,45.73)(94.66,44.93)
\bezier{22}(94.66,44.93)(91.44,44.23)(88.39,44.93)
\bezier{40}(116.60,69.99)(116.60,72.30)(108.76,73.45)
\bezier{40}(108.76,73.45)(100.93,74.61)(93.09,73.45)
\bezier{40}(93.09,73.45)(85.26,72.30)(85.26,69.99)
\bezier{40}(116.60,69.99)(116.60,67.68)(108.76,66.53)
\bezier{40}(108.76,66.53)(100.93,65.37)(93.09,66.53)
\bezier{40}(93.09,66.53)(85.26,67.68)(85.26,69.99)
\put(85.26,14.31){\line(0,1){55.68}}
\put(116.60,14.31){\line(0,1){55.68}}
\thicklines
\multiput(88.78,67.48)(0.50,0.12){46}{\line(1,0){0.50}}
\multiput(88.39,10.96)(0.42,0.12){58}{\line(1,0){0.42}}
\bezier{260}(88.78,67.48)(122.87,36.21)(88.39,10.96)
\bezier{74}(111.90,72.77)(117.38,64.60)(116.60,53.28)
\bezier{64}(115.03,31.57)(111.90,39.55)(115.03,47.72)
\bezier{40}(115.03,47.72)(116.60,53.28)(116.60,53.28)
\bezier{44}(115.03,31.57)(116.60,27.11)(116.60,25.44)
\bezier{22}(116.60,25.44)(116.60,20.99)(112.68,17.93)
\bezier{17}(111.90,72.77)(103.02,65.16)(98.58,56.63)
\bezier{114}(98.58,56.63)(91.53,42.89)(103.28,28.23)
\bezier{12}(103.28,28.23)(106.41,23.40)(112.68,17.93)
\bezier{74}(88.39,10.96)(84.47,19.69)(85.26,31.01)
\bezier{64}(86.82,53.84)(89.96,44.75)(86.82,36.58)
\bezier{40}(86.82,36.58)(85.26,31.01)(85.26,31.01)
\bezier{44}(86.82,53.84)(85.26,57.18)(85.26,59.97)
\bezier{22}(85.26,60.52)(85.26,63.31)(88.78,67.48)
\thinlines
\put(87.61,67.02){\vector(4,1){0.2}}
\multiput(78.99,64.98)(0.51,0.12){17}{\line(1,0){0.51}}
\put(78.21,65.53){\makebox(0,0)[cc]{$E$}}
\put(117.38,42.15){\makebox(0,0)[lc]{$t=0$}}
\bezier{40}(116.60,14.31)(116.60,16.25)(112.68,17.93)
\bezier{10}(112.68,17.93)(104.85,20.43)(98.58,19.87)
\bezier{8}(98.58,19.87)(90.35,19.32)(86.43,16.53)
\bezier{10}(86.43,16.53)(85.26,15.59)(85.26,14.31)
\bezier{16}(85.26,42.15)(84.94,43.71)(88.39,44.97)
\bezier{5}(88.39,44.97)(91.94,46.23)(95.91,46.60)
\bezier{40}(95.91,46.60)(99.89,46.98)(103.96,46.75)
\bezier{7}(103.96,46.75)(110.02,46.31)(114.09,44.67)
\bezier{12}(114.09,44.67)(116.18,44.00)(116.60,42.15)
\put(28.00,5.00){\makebox(0,0)[cc]{(a)}}
\put(101.00,5.00){\makebox(0,0)[cc]{(b)}}
\bezier{92}(89.00,67.00)(88.61,57.78)(94.00,45.00)
\bezier{140}(89.00,67.00)(103.89,56.39)(108.00,40.00)
\put(94.00,44.80){\line(3,-1){11.50}}
\put(93.00,48.00){\line(3,-1){12.00}}
\put(92.00,51.00){\line(3,-1){12.00}}
\put(91.00,54.00){\line(3,-1){11.80}}
\put(90.00,57.00){\line(3,-1){11.00}}
\put(89.50,60.00){\line(3,-1){9.00}}
\put(89.00,62.50){\line(3,-1){7.00}}
\multiput(89,65)(0.42,-0.13){8}{\line(1,0){0.42}}
\bezier{5}(105.4,41.00)(106.5,40.50)(107.99,40.00)
\bezier{3}(105.3,44.00)(106.1,43.65)(106.94,43.30)
\bezier{3}(103.94,47.00)(104.9,46.75)(105.78,46.50)
\bezier{3}(102.8,50.00)(103.65,49.75)(104.5,49.50)
\bezier{2}(101.28,53.17)(101.7,53.08)(102.11,53.00)
\bezier{2}(98.44,56.83)(99.22,56.66)(99.94,56.50)
\bezier{2}(96.44,60.00)(97.72,59.91)(97.00,59.82)
\put(30.39,18.00){\vector(1,2){3.22}}
\put(27.94,18){\makebox(0,0)[ct]{$\rho=0$}}
\put(27.94,14.89){\makebox(0,0)[ct]{or $P=\ell\sqrt{m}$}}
\bezier{144}(108.00,40.00)(117.78,61.11)(112.00,73.00)
\end{picture}

\caption{Time development of a single black hole without angular momentum.
(a) Isometric embedding of the subspace $\phi = 0$ or the subspace $T=0$
as a 1+1-dimensional adS space in 2+1-dimensional flat space.
(b) Representation of three-dimensional adS spacetime as the interior of 
a cylinder in ``sausage coordinates.'' A black hole spacetime is the double 
of the heavily outlined region. The striped surface is the horizon of the
left front null infinity, whose endpoint is $P$.}
\end{figure}
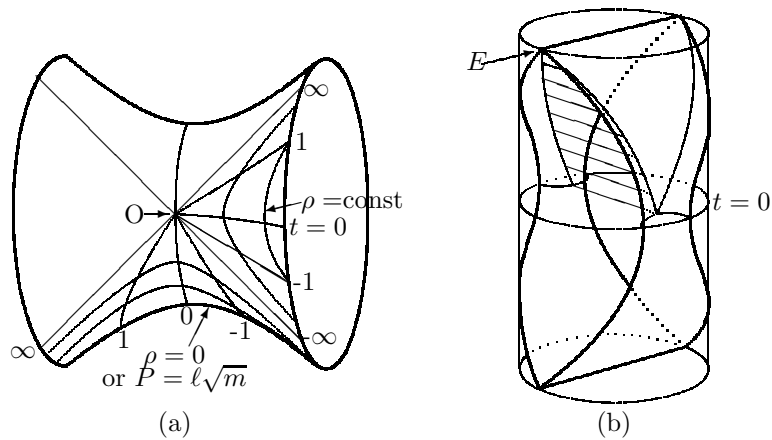

Expressed in new coordinates, 
\begin{equation}
P^2 = \ell^2m-\rho^2, \qquad T=\ell\phi, \qquad \Phi=t/\ell
\end{equation}
the metric (\ref{BTZ}) is the same expression as in the old coordinates.
The new coordinates interchange the role of $t$ and $\phi$
in the region $\rho$ or $P<\ell\sqrt{m}$, where these coordinates are
timelike, and the spacelike surfaces are analogous to a Kantowski-Sachs 
universe. Therefore Fig.\ 3a can
also be regarded as a picture of the $t=0$ section of
(\ref{BTZ}). In that case the numbers label the values of $\phi$, and 
the initial state is the curve labeled $P = \ell\sqrt{m}$.

To return to the interpretation of (\ref{BTZ})
as a black hole, we make $\phi$ periodic by identifying $\phi=-\pi$
and $\phi=+\pi$. Then all $P$ or $\rho=$ const curves are circles, 
with $\rho=0$ the throat of the ``wormhole'' geometry, which collapses 
to zero size at O as the timelike $\rho$ increases. The origin O, which 
previously was a coordinate singularity, becomes a non-Hausdorff singularity 
at $P=0$, as in Misner space \cite{Mis}. This singular line extends to 
infinity and marks the endpoint $E$ of null infinity. Therefore the past 
of null infinity has a boundary, the horizon of the black hole.

Figure 3b is a representation adS spacetime as the interior of a cylinder
in ``sausage coordinates'' \cite{Ingemar}. Each horizontal slice of 
the cylinder is a Poincar\'e disk as in Fig.\ 1b, and the time coordinate is 
one in which the adS space appears static. The mantle of the cylinder
represents infinity. The heavily outlined region is half ($\phi = 0$ to
$\phi = \pi$, for example) of the BTZ spacetime (\ref{BTZ}). After doubling, 
the spacetime is no longer static, for example because the boundaries 
where the gluing takes place approach each other and intersect in a geodesic 
that ends at the point $E$ at infinity. The heavily-outlined lozenge-shaped 
regions on the left front and right rear of the cylinder are the two 
{\scri}s. The one in front has endpoint at $E$. The horizon (striped surface) 
is the backward lightcone from $E$.

\subsection{Time development of multi-black-holes}
Each exterior region of a multi-black-hole initial state is isometric to a 
BTZ exterior, therefore the time development of each exterior will also be
isometric to that shown in Fig.\ 3b. In particular, as seen from one such
exterior, the other black holes lie behind that exterior's horizon. The 
whole spacetime up to the non-Hausdorff singularity can be obtained by 
doubling the regions of adS space corresponding to the initial neighborhoods 
(such as those of Fig.\ 2a or c). The boundaries (``seams'') of these 
spacetime regions are generated by timelike geodesics normal to the initial 
boundaries. 

Such two-dimensional, timelike boundaries are totally geodesic, and 
therefore fit together smoothly. Their intrinsic geometry is constant 
negative curvature (two-dimensional adS). The normal geodesics do not 
generate a complete surface, but only the part of a two-dimensional adS 
space that lies in the domain of dependence of the initial surface. Because 
all normal geodesics to a time-symmetric surface in adS spacetime intersect 
in one point, all the seams also intersect in one point $T$. When they 
are analytically extended to complete surfaces they intersect along 
spacelike geodesics, which will form the non-Hausdorff singularity after 
gluing. Thus the top (and bottom) of the region to be doubled looks like a 
pyramid-shaped ``tent'' whose ridge lines are the singularities (Fig.\ 4a).
In the interior the ridge lines come together at the point $T$. From there
they run to infinity, where they define the endpoints of each exterior's \scri. The 
horizon is obtained by running a lightcone backwards from each of these 
endpoints to the points of intersection with another such backward lightcone. 
(For details of this construction see \cite{DBS}.)

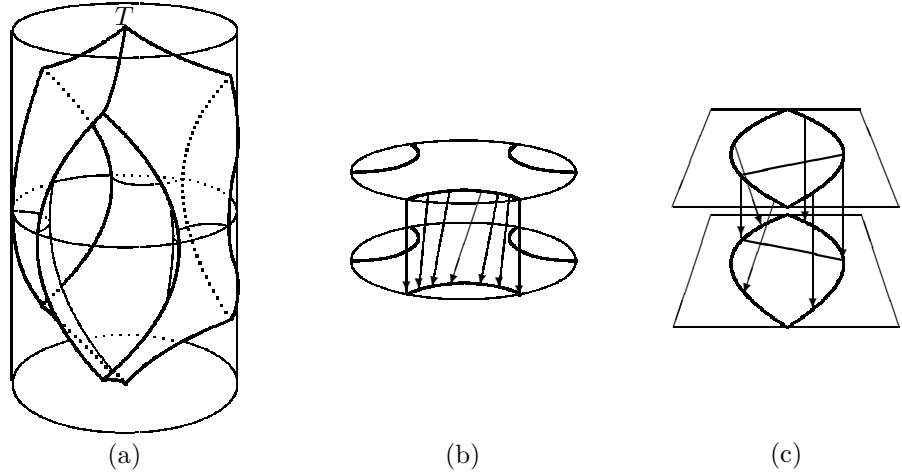
\begin{figure}
\unitlength 0.50mm
\linethickness{0.4pt}
\begin{picture}(246.00,122.16)
\bezier{50}(54.92,57.50)(40.00,55.07)(25.08,57.50)
\bezier{54}(25.08,57.50)(10.15,60.58)(10.15,66.00)
\thicklines
\bezier{228}(34.00,21.00)(74.50,55.33)(34.00,92.00)
\bezier{200}(34.00,92.00)(-1.00,57.33)(34.00,21.00)
\bezier{34}(68.00,101.70)(70.00,93.00)(70.00,85.00)
\bezier{16}(70.00,85.00)(70.00,81.00)(69.00,77.00)
\bezier{30}(69.00,77.00)(67.50,70.33)(67.50,62.50)
\bezier{34}(68.00,37.00)(70.00,40.00)(70.00,46.00)
\bezier{16}(70.00,46.00)(70.00,49.00)(69.00,52.00)
\bezier{30}(69.00,52.00)(67.50,56.00)(67.50,61.50)
\bezier{144}(18.00,41.00)(2.00,59.33)(18.00,104.00)
\bezier{3}(19.50,42.50)(18.75,41.75)(18.00,41.00)
\bezier{40}(68.00,37.00)(42.33,76.33)(68.00,101.70)
\bezier{10}(31.00,89.00)(25.33,97.33)(18.00,104.00)
\bezier{64}(68.00,102.00)(50.00,107.33)(40.00,115.00)
\bezier{50}(40.00,115.00)(29.00,106.33)(18.00,104.00)
\bezier{48}(40.00,115.00)(37.00,96.33)(34.00,92.00)
\thinlines
\put(40.00,118.00){\makebox(0,0)[cc]{$T$}}
\bezier{54}(69.85,114.00)(69.85,118.05)(54.92,120.35)
\bezier{50}(54.92,120.35)(40.00,122.16)(25.08,120.35)
\bezier{54}(25.08,120.35)(10.15,118.05)(10.15,114.00)
\bezier{54}(69.85,114.00)(69.85,109.95)(54.92,107.65)
\bezier{50}(54.92,107.65)(40.00,105.84)(25.08,107.65)
\bezier{54}(25.08,107.65)(10.15,109.95)(10.15,114.00)
\put(70.00,66.00){\line(0,1){48.00}}
\put(10.00,66.00){\line(0,1){48.00}}
\bezier{54}(69.85,20.00)(69.85,27.27)(54.92,31.40)
\bezier{54}(25.08,31.40)(10.15,27.27)(10.15,20.00)
\bezier{54}(69.85,20.00)(69.85,12.73)(54.92,8.60)
\bezier{50}(54.92,8.60)(40.00,5.35)(25.08,8.60)
\bezier{54}(25.08,8.60)(10.15,12.73)(10.15,20.00)
\thicklines
\bezier{14}(40.00,20.00)(39.00,21.33)(34.00,21.00)
\bezier{66}(68.00,37.00)(48.00,29.33)(40.00,20.00)
\bezier{3}(40.17,19.83)(38.60,21.70)(37.00,23.67)
\thinlines
\bezier{13}(55.00,31.00)(39.33,35.00)(26.00,31.00)
\put(70.00,66.00){\line(0,-1){46.00}}
\put(10.00,66.00){\line(0,-1){45.00}}
\bezier{9}(55.00,74.56)(44.56,75.50)(35.89,75.44)
\bezier{32}(35.89,75.44)(27.22,75.44)(20.33,73.44)
\bezier{8}(20.33,73.44)(9.44,69.89)(10.11,66.11)
\bezier{68}(54.33,57.44)(75.00,61.44)(67.89,69.89)
\bezier{7}(67.89,69.89)(63.00,73.22)(55.22,74.56)
\bezier{126}(36.58,23.55)(15.67,48.09)(21.12,74.76)
\thicklines
\bezier{114}(31.12,89.00)(44.76,68.70)(22.94,45.97)
\bezier{3}(22.94,45.97)(21.73,44.45)(19.30,42.64)
\thinlines
\bezier{46}(51.39,43.72)(53.61,56.50)(52.22,65.94)
\bezier{10}(16.44,60.11)(20.00,61.44)(20.00,62.78)
\bezier{3}(20.00,62.78)(20.00,63.89)(17.56,65.00)
\bezier{16}(17.56,65.00)(15.78,65.89)(10.00,65.89)
\bezier{3}(55.56,74.56)(52.67,73.44)(49.56,73.22)
\bezier{30}(49.56,73.22)(38.89,72.11)(36.22,75.22)
\thicklines
\bezier{10}(22.08,36.92)(30.00,29.63)(36.04,23.17)
\bezier{24}(17.67,40.67)(20.00,39.33)(22.33,36.33)
\thinlines
\bezier{26}(67.50,61.92)(60.83,63.58)(54.17,60.88)
\bezier{6}(54.37,57.54)(52.50,58.58)(52.92,59.42)
\put(40.00,0.67){\makebox(0,0)[cc]{(a)}}
\bezier{54}(100.15,52.69)(100.15,58.38)(115.08,61.62)
\bezier{50}(115.08,61.62)(130.00,64.17)(144.92,61.62)
\bezier{54}(144.92,61.62)(159.85,58.38)(159.85,52.69)
\bezier{54}(100.15,52.69)(100.15,46.99)(115.08,43.76)
\bezier{50}(115.08,43.76)(130.00,41.21)(144.92,43.76)
\bezier{54}(144.92,43.76)(159.85,46.99)(159.85,52.69)
\thicklines
\bezier{38}(115.00,43.80)(119.67,46.31)(130.00,46.88)
\bezier{38}(130.00,46.88)(140.67,46.31)(145.00,43.80)
\bezier{40}(144.00,61.58)(139.67,58.50)(145.00,55.42)
\bezier{40}(145.00,55.42)(149.67,52.69)(160.00,52.69)
\bezier{40}(116.00,61.58)(120.33,58.50)(115.00,55.42)
\bezier{40}(115.00,55.42)(110.33,52.69)(100.00,52.69)
\thinlines
\bezier{54}(100.15,76.33)(100.15,81.02)(115.08,83.68)
\bezier{50}(115.08,83.68)(130.00,85.78)(144.92,83.68)
\bezier{54}(144.92,83.68)(159.85,81.02)(159.85,76.33)
\bezier{54}(100.15,76.33)(100.15,71.65)(115.08,68.99)
\bezier{50}(115.08,68.99)(130.00,66.89)(144.92,68.99)
\bezier{54}(144.92,68.99)(159.85,71.65)(159.85,76.33)
\thicklines
\bezier{38}(115.00,69.02)(119.67,71.08)(130.00,71.55)
\bezier{38}(130.00,71.55)(140.67,71.08)(145.00,69.02)
\bezier{40}(144.00,83.65)(139.67,81.11)(145.00,78.58)
\bezier{40}(145.00,78.58)(149.67,76.33)(160.00,76.33)
\bezier{40}(116.00,83.65)(120.33,81.11)(115.00,78.58)
\bezier{40}(115.00,78.58)(110.33,76.33)(100.00,76.33)
\thinlines
\put(135.00,71.33){\vector(-1,-3){8.33}}
\put(115.00,69.33){\vector(0,-1){25.00}}
\put(145.00,69.33){\vector(0,-1){26.00}}
\put(118.33,45.33){\vector(0,-1){0.2}}
\multiput(120.67,70.33)(-0.12,-1.25){20}{\line(0,-1){1.25}}
\put(139.67,45.67){\vector(0,-1){0.2}}
\multiput(142.33,70.00)(-0.12,-1.06){23}{\line(0,-1){1.06}}
\put(134.67,47.00){\vector(-1,-4){0.2}}
\multiput(139.33,70.67)(-0.12,-0.61){39}{\line(0,-1){0.61}}
\put(130.00,0.67){\makebox(0,0)[cc]{(b)}}
\put(122.00,46.67){\vector(-1,-4){0.2}}
\multiput(126.33,71.00)(-0.12,-0.66){37}{\line(0,-1){0.66}}
\put(186.00,67.00){\line(1,0){60.00}}
\put(246.00,67.00){\line(-2,5){10.40}}
\put(235.60,93.00){\line(-1,0){39.60}}
\put(196.00,93.00){\line(-2,-5){10.40}}
\put(196.00,65.00){\line(-1,-3){10.00}}
\put(186.00,35.00){\line(1,0){60.00}}
\put(246.00,35.00){\line(-1,3){10.00}}
\put(236.00,65.00){\line(-1,0){40.00}}
\thicklines
\bezier{120}(216.00,93.00)(186.00,82.00)(216.00,67.00)
\bezier{120}(216.00,67.00)(246.00,82.00)(216.00,93.00)
\bezier{120}(216.00,65.00)(186.00,54.00)(216.00,35.00)
\bezier{120}(216.00,35.00)(246.00,54.00)(216.00,65.00)
\put(231.00,81.00){\line(-5,-1){27.00}}
\put(231.00,53.00){\line(-5,1){27.00}}
\thinlines
\put(204.00,76.00){\vector(0,-1){17.00}}
\put(231.00,81.00){\vector(0,-1){28.00}}
\put(213.00,69.00){\vector(-1,-3){8.33}}
\put(202.00,84.50){\vector(1,-3){7.5}}
\put(223.00,71.00){\vector(0,-1){31.00}}
\put(221.00,91.00){\vector(0,-1){28.00}}
\put(216.00,1.00){\makebox(0,0)[cc]{(c)}}
\end{picture}
\caption{(a) The time development in sausage coordinates of a region of
adS spacetime that becomes a three-black-hole when doubled. (b) An alternative
way of gluing two regions of Poincar\'e disks, seen in perspective, to
obtain a three-black-hole initial state. (c) An alternative way of gluing
two vertical (timelike) slices of the ``sausage'' of Fig.\ 3.}
\end{figure}

\section{Angular momentum}
The general BTZ metric describes a ``single'' black hole (with two asymptotically
adS regions) that has angular momentum $J$ in addition to mass $m$. As we will
see below, the metric with $J\neq 0$ can be obtained from the time-symmetric
one, which has $J=0$, by changing the rules by which its two halves are glued
together. To fix ideas we first consider such rule change within the 
time-symmetric class. 

\subsection{Alternative ways of gluing}
Consider a three-black-hole initial state, obtained by gluing together two
copies of the region between three disjoint geodesics on the Poincar\'e disk. 
Previously we have identified each point of the heavily drawn curves in the
upper disk of Fig.\ 4b with the one vertically below it on the lower disk.
The geodesic's neighborhood is invariant under ``translation" isometries
that move each point on the geodesic by a constant distance. Therefore we
get an equally smooth surface if we move the points on the lower disk by
such an isometry, so that the gluing identifies points that are connected
by the arrows in Fig.\ 4b. The result of gluing with a shift depends on the
amount of shift, for example because the size of the minimal closed geodesics 
around the adjacent black holes, and hence their masses, depend on it. 
However, a change in mass is all that can happen to the initial state,
because we know that all time-symmetric three-black-hole initial states
are characterized by just three mass parameters. The shift can always be 
``transformed away'' by changing the geodesic seams that are to be glued 
together.\footnote{The same circumstance in flat space is illustrated by gluing a
cylinder out of a piece of paper either with or without such a shift. 
In either case one gets a cylinder. If there is no shift, the seam is
parallel to the cylinder's axis. If there is a shift, the cylinder's radius 
is smaller, and the seam is a helix on the cylinder.}

In the space-time picture of a black hole (Fig.\ 3b) or of multi-black-holes
(Fig.\ 4a) the seams are timelike hypersurfaces of constant negative curvature. These
surfaces are invariant under a 3-parameter group of isometries. Again we
can consider different gluing rules, depending on what isometry is
applied at the seam. We want to distinguish those ways of re-gluing 
time-symmetric black holes that lead to new types of spacetimes. 

We have already seen that we do not get a new class of spacetimes 
from re-gluing a seam by an isometry that leaves the surface of
time-symmetry invariant. Similarly, if we re-glue a time-symmetric 
BTZ black hole by a time translation, we again get a spacetime with
a surface of time-symmetry, that is, another time-symmetric black hole.
This is illustrated in Fig.\ 4c in the timelike subspace obtained by
slicing Fig.\ 3b with a vertical plane from left back to right front. 
Two copies of this plane are shown in perspective. The two pairs of curves
on the planes are the seams, at $\phi=0$ and $\pi$ on the top plane,
and $\phi=\pi$ and $2\pi$ on the bottom plane. The arrows connect points
that are to be glued together.\footnote{This alternative to gluing 
along vertical arrows can also be illustrated in Fig.\ 3a, where the
numbers at the bottom are values of $\phi$: cut the figure into two halves
by a vertical plane through the center and perpendicular to the picture
plane, rotate one half against the other about a horizontal axis, and
re-glue.} The heavy lines on the planes connect smoothly
to form a closed geodesic\footnote{In an accurate plot of sausage 
coordinates these geodesic segments would not look straight as they do in this
qualitative picture.} in the new surface of time-symmetry produced by 
this re-gluing. (In the three-dimensional picture the new surface of time-symmetry
is obtained from the old one (labeled $t=0$ in Fig.\ 3b) by a ``Lorentz 
transformation" isometry that has the geodesic $t=0,\, \phi=\pi$ as an axis.)

In order to obtain a new class of black holes, with angular momentum, we
re-glue a time-symmetric black hole by an isometry in the seam that has a 
fixed point at $t=0$.

\subsection{BTZ black hole with angular momentum}
In the metric (\ref{BTZ}) for the static BTZ black hole, introduce new
coordinates $T,\,\varphi,\,R$,
\begin{eqnarray}
t = T + \left({J\over 2m}\right)\varphi\nonumber \\
\phi = \varphi + \left({J\over 2m\ell^2}\right)T \\
R^2=\rho^2\left(1-{J^2\over 4m^2\ell^2}\right) + {J^2\over 4m}\nonumber
\label{coord}
\end{eqnarray}
where $J<2m\ell$ is a constant with dimension of length, and define another new
constant
\begin{equation}
M = m + {J^2\over 4m\ell^2}.
\label{mass}
\end{equation}
In terms of these new quantities the metric (\ref{BTZ}) becomes
\begin{equation}
ds^2 = -N^2 dT^2 + N^{-2}dR^2 + R^2\left(d\varphi + {J\over 2R^2}
dT\right)^2
\label{BTZJ}
\end{equation}
where
\begin{equation}
N^2 = \left({R\over\ell}\right)^2 - M + \left({J\over 2R}\right)^2.
\end{equation}
Equation (\ref{BTZJ}) is the metric for a black hole with angular momentum $J$.
In this metric, the new coordinate $\varphi$ is taken as periodic. When it 
changes by its period, $2\pi$, the old coordinates of (\ref{BTZ}) change
by
\begin{equation}
t \rightarrow t + {\pi J\over m} \qquad \phi \rightarrow \phi + 2\pi
\end{equation}
This tells us that in order to obtain a black hole with angular momentum,
by re-gluing a time-symmetric one, we should apply a ``boost'' by $\pi J/m$
at the $\phi = 2\pi$ seam about the (old) horizon. 

The construction by re-gluing a $J=0$ black hole spacetime does not yield
the full $J\neq 0$ spacetime, because the pieces that we glue together
end at the ridge lines ($r=0$) that become non-Hausdorff singularities in the
$J=0$ case. When $J\neq 0$ the spacetime can be extended beyond the
ridge line, to $R = 0$, which would correspond to negative $r^2$. Otherwise
stated, we do not obtain a representation of the full spacetime when we
cut it into two pieces, because the two cuts we make along the seams
intersect each other when we get too far from the initial surface. Nevertheless,
a piece of the spacetime is enough to characterize it, and we can use it
to deduce the number of parameters needed.

\subsection{Multi-black-holes with angular momenta}
Our purpose is to characterize spacetimes that have angular momenta and
are counterparts to the
time-symmetric multi-black-hole spacetimes of section 2. The basic building
block is the three-black-hole spacetime of Fig.\ 4a. It has three totally
geodesic seams, each of which is a 1+1-dimensional adS spacetime, of constant
negative curvature $\Lambda$, exactly the same as the seams the single black hole of
section 3.2. Each of these seams can therefore be re-glued smoothly by
applying an adS isometry. These isometries form a three-parameter group.
As before, only a one-dimensional subset leads to spacetimes without any
surface of time-symmetry, so there is one effective parameter per seam.
The general three-black-hole state is therefore characterized by three
configuration parameters, which can be taken to be the three masses,
and three momentum parameters, the boost angles at the three seams.

The actual angular momentum of any one of the black holes (and therefore
also its actual mass $M$, equation (\ref{mass})) depends on the boost parameters 
of its two adjacent seams, and on the fixed point of the boost. 
Note that a boost at a seam is an isometry only within that seam (and in a neighborhood  
of the seam), but it cannot in general be extended to the whole spacetime. 
Once we have a three-black-hole spacetime with angular momentum, we can forget
about the time-symmetric geometry that was used to construct it. The core
will extend only to each leg's local (outer) horizon at $R_{\rm H}$ 
(where $R_{rm H}$ is the larger root of $N^2(R_{\rm H})=0$). 
The geometry is locally reflection-symmetric about $R_{\rm H}$, 
and it can be matched there to other cores with the same local geometry.

The general time-symmetric multi-black-hole spacetime considered here was
put together out of three-black-hole cores and and exteriors (Fig.\ 2b).
We can put a spacetime together in topologically the same way if the cores
have angular momenta. We only have to match masses and angular momenta
at the seams, because the neighborhood of each seam (including the entire 
interior region, $R < R_{\rm H}$)
depends only on that seam's $M$ and $J$. So, when matching two cores we
lose two of the parameters characterizing the separate cores. In the
time-symmetric case we also gained one parameter, which we called a twist
because it describes a rotation along the seam. In the case of space-time,
the neighborhood of a seam is the same as a single black hole metric (\ref{BTZJ}).
It is therefore locally invariant no only under a change in $\phi$ (``twist'') 
but also under a change in $T$ {``boost''}. Re-identifying an internal
seam with a twist and a boost gives us two additional parameters back.
Thus our general multi-black-hole is
characterized by twice the number of parameters necessary to specify a
time-symmetric one.

\section{Conclusions}
We have seen that sourceless 2+1-dimensional Einstein
theory with a negative cosmological constant admits solutions with all spatial
two-dimensional topologies that can carry a constant negative curvature
metric.\footnote{In addition, the torus topology is also admitted, but not
as a time-symmetric state. For example, identifying $t$ and $\phi$
periodically for $\rho < \ell\sqrt{m}$ in (\ref{BTZ}) yields the torus
topology.} Among these are multi-black-hole spacetimes that have several
asymptotically anti-de Sitter regions, each of which is characterized by
a mass and an angular momentum. A number of additional parameters are needed
(except for the three-black-hole configuration) to characterize the internal
structure. Half of these are configuration parameters that specify internal
sizes or angular relationships; the other half are momentum parameters,
which vanish if the space-time is time-symmetric.

\end{document}